\documentclass[aps,prd,preprintnumbers, nofootinbib, onecolumn]{revtex4-1}
\usepackage{bm}
\usepackage{latexsym}
\usepackage{amsmath,amsfonts,amssymb}
\usepackage{graphicx,epsfig}
\usepackage{psfrag}
\usepackage{amsthm}
\usepackage{color}

\usepackage{amsmath}
\usepackage{braket}
\usepackage{indent first}
\usepackage[normalem]{ulem}
\useunder{\uline}{\ul}{}
 
\DeclareMathOperator{\arcsinh}{arcsinh}
\begin{document}
\title{{\bf Gravitation as a source of decoherence}}
\author{Saurya Das}
\email{saurya.das@uleth.ca}
\author{Matthew P. G. Robbins}
\email{matthew.robbins@uleth.ca}
\affiliation{Theoretical Physics Group and Quantum Alberta, Department of Physics and Astronomy, University of Lethbridge, 4401 University Drive, Lethbridge, Alberta T1K 3M4, Canada}
\author{Elias C. Vagenas}
\email{elias.vagenas@ku.edu.kw}
\affiliation{Theoretical Physics Group, Department of Physics, Kuwait University, P.O. Box 5969, Safat 13060, Kuwait}
 
\begin{abstract}
\par\noindent
It is believed that classical behavior emerges in a quantum system due to decoherence.
It has also been proposed that gravity can be a source of this decoherence.
We examine this in detail by studying a number of quantum systems, including ultra-relativistic and
non-relativistic particles, at low and high temperatures in an expanding Universe, and show that this proposal
is valid for a large class of quantum systems.
 
\end{abstract}
 
\maketitle
 
\section{Introduction}
%
\par\noindent
An outstanding question in theoretical physics is how classical behavior emerges
from a quantum mechanical
framework. It is believed that interactions between a system and its environment cause
the system to lose its ability to form a coherent superposition, in a process known as decoherence \cite{Zurek:2003(1)zz, Schlosshauer}.
One proposed mechanism for the Universe exhibiting observed
classical behavior is decoherence via gravitation
%
\cite{Hu:2014kia, Anastopoulos:2013zya,Blencowe:2012mp,DeLorenci:2014vwa}.
In this paper, we study this possibility, focusing on
the recent work of Refs. \cite{Blencowe:2012mp} and \cite{DeLorenci:2014vwa}, 
in which the decoherence effects of gravitons on quantum systems at high
and low temperatures were studied.
We show that in several cases, gravity can indeed be the source of decoherence in our Universe.
\par\indent
Following \cite{Blencowe:2012mp} and \cite{DeLorenci:2014vwa},
we consider a thermal bath of gravitons in a Minkowski background (with small deviations).
The gravitationally induced decoherence time for a non-relativistic macroscopic system in the {\it high temperature}
limit
(under the Born-Markov approximation) is given by \cite{Blencowe:2012mp}
\begin{equation}
t_d=\frac{\hbar}{k_BT}\left(\frac{E_P}{\Delta\omega}\right)^2\,
\label{eq: Blencowe}
\end{equation}
where $k_B$ is Boltzmann's constant, $T$ is the temperature of the bath, $\Delta\omega$ is the difference
between the energy levels of two states in a superposition, and $E_P$ is  the Planck energy.
At this point, it should be stressed that
although Ref.\cite{Blencowe:2012mp} starts off with the non-relativistic
energy-momentum dispersion relation, the rest of the derivation is
done using relativistic quantum field theoretic methods. Therefore
Eq.(\ref{eq: Blencowe}) can rightly be applied to relativistic systems
(as well as their non-relativistic limits).
%
%
 
\par\indent
On the other hand, also for gravitationally
induced decoherence of non-relativistic systems of arbitrary size, due to
Gaussian metric fluctuations around flat spacetime, and at {\it low temperatures}, the
decoherence time was estimated to be \cite{DeLorenci:2014vwa}
\begin{equation}
t_d=\frac{1}{4}\sqrt{\frac{27}{\pi}}\frac{E_P\hbar}{k_BT\Delta\omega}\,.
\label{eq: Ford 1}
\end{equation}
\par\noindent
Furthermore, if we consider a bath of gravitons of energy density $\rho_g$,
then the decoherence time, again in the low-temperature limit, is given by \cite{DeLorenci:2014vwa}
\begin{equation}
t_d=\sqrt{\frac{9}{32\pi}}\frac{E_P}{\lambda_g\Delta\omega}\sqrt{\frac{\hbar}{\rho_gc}}
\label{eq: Ford 2}
\end{equation}
where $\lambda_g$ is the graviton wavelength.
We point out that although 
Eqs.(\ref{eq: Ford 1}) and (\ref{eq: Ford 2}) 
were initially proven for non-relativistic systems, they can
also be applied for relativistic systems. The reason is that
(as shown in Refs. \cite{Alberto:1996, Alberto:2011wn})
the wavefunctions used for non-relativistic particles are the same
(at least for the first energy eigenstates)
with those used for relativistic particles and the energy formula for the particle in a
box depends only on the boundary conditions $\Psi(0)=\Psi(L)=0$ with $L$ to be the size of the box.
The criterion to decide if a particle in a box is non-relativistic or relativistic is the size $L$ of the box.
Thus, the derivations in Ref. \cite{DeLorenci:2014vwa} goes through for relativistic particles and at the end one has to substitute the corresponding energy formula. We do precisely this in this paper for the relevant cases.
\par\noindent
\par\indent
We will consider the above decoherence times for a variety of quantum systems
in an expanding Universe, and check if these times are indeed less than the current age
of the Universe ($\approx 4 \times 10^{17}~$ s).
In such cases, one would be able to conclude that
gravitation had sufficient time to decohere quantum systems, and in fact may have been the
primary source of decoherence.
On the other hand, decoherence times exceeding the
Universe's age would imply that there was not enough time for gravitational decoherence to occur and subsequent classical behaviour to emerge. Consequently, this
would rule out the proposed models.

\par\indent
To determine the age of the Universe $t_U$
in a give epoch characterized by the scale factor $a$,
or equivalently the red-shift $z$,
we employ the following relations, valid for early and late epochs, 
respectively \cite{Carmeli:2005if} and  \cite{Macdonald:2006sa},
(more on this in  the Discussion section; we assume $a=1$ in the current epoch)
\begin{eqnarray}
&& t_U\approx\frac{28\pi\times10^{16}}{1+1/a^2}~,
\label{eq: carmeli} \\
&& t_U=\frac{2H_0^{-1}}{3\Omega_{\Lambda}^{1/2}}\arcsinh\left[\left(\frac{\Omega_{\Lambda}}
{\Omega_M}\right)^{1/2}a^{3/2}\right]
\label{eq: macdonald}
\end{eqnarray}
where $H_0=2.19\times10^{-18}\text{ s$^{-1}$}$ is the Hubble constant, $\Omega_{\Lambda}=0.689$ is
the dark energy density, and $\Omega_M=0.311$ is the matter density
\footnote{All values are averaged
from the data released by the Planck mission in 2015 \cite{Ade:2015xua}.}.
We also use the following in our calculations:
$z=1/a-1, T=T_0/a, L=L_0 a$,
where $T_0=2.73$~K is the current CMBR temperature, and
$L_0 \approx 8.8 \times 10^{26}$~m
and $L$ are the Hubble radius at the current epoch
and at any given epoch, respectively.
Note that given the spatial flatness of the Universe, and the
existence of a horizon at any epoch of extent $\approx L_0 a$,
it is reasonable to approximate a quantum system, relativistic or non-relativistic,
as a particle in a very large box (of length $L_0a$).
We also take $\Delta\omega$ that appears in
Eqs.(\ref{eq: Blencowe}),
(\ref{eq: Ford 1}), and (\ref{eq: Ford 2}) to be the energy difference between the
ground state and the first excited state.
\par\indent
Now, we consider a number of quantum systems, characterized by the following values of $\Delta\omega$. The examples we consider, namely ultra-relativistic
and non-relativistic particles,
gravitons and atomic systems, although are not
exhaustive, cover most realistic scenarios and the matter content of our Universe
:\\
 
\noindent
(i) \underline{Ultrarelativistic particles} \\
\begin{equation}
\Delta\omega=\frac{\pi\hbar c}{L_{0}a}
\label{energy1}
\end{equation}
Note that the above being mass independent, holds for
any ultra-relativistic particle, including for example electrons and gravitons.

\vspace{0.1cm}
\noindent
(ii) \underline{Non-relativistic electrons ($m=0.5~MeV/c^2$)} \\
\begin{equation}
\Delta\omega=\frac{\pi^2\hbar^2}{m(L_0a)^2}
\label{energy2}
\end{equation}

\vspace{0.1cm}
\noindent
(iii) \underline{Non-relativistic gravitons ($m_g=10^{-32}~eV/c^2$)} \\
\begin{equation}
\Delta\omega=\frac{\pi^2\hbar^2}{m_g(L_0a)^2}
\label{energy3}
\end{equation}
As was shown in \cite{Das:2014agf}, non-relativistic gravitons can form part of
a {\it macrosopic} Bose-Einstein condensate at very early epochs, due to
a very high critical temperature of the condensate.
Note that the decoherence time for these gravitons can be estimated by using
Eqs.(\ref{eq: Blencowe}) and (\ref{eq: Ford 1}),
applicable to non-relativistic systems, as well as by using Eq.(\ref{eq: Ford 2}),
also applicable to non-relativistic systems and requiring a background bath of gravitons.

\vspace{0.2cm}
\noindent
(iv) \underline{Atomic systems}
 
We use a characteristic value of
$\Delta\omega\approx1$ eV \cite{Blencowe:2012mp}.

\vspace{0.2cm}
\par\indent
The rest of the paper is organized as follows.
In Tables I, II and III, we compute the decoherence times
$t_d$ by equating (\ref{eq: Blencowe}), (\ref{eq: Ford 1}), and (\ref{eq: Ford 2}),
respectively, to the Universe's age estimated by Eq.(\ref{eq: carmeli}).
 
In Tables IV, V and VI,
we repeat the calculations using Eq.(\ref{eq: macdonald})
for estimating the age of the Universe.
When applying equation (\ref{eq: Ford 2}),
we assume  $\rho_g$ to be equal to the energy density of dark matter, i.e.,
$\rho_{DM}=0.25\rho_{crit}c^2/a^3$, as
discussed in Ref. \cite{Das:2014agf}.
We also present the corresponding values of $L_0a$ and $z$ in the above Tables.

In particular, when we  substitute into the expression for the decoherence time, i.e.,   
Eq. (\ref{eq: Blencowe}), or Eq. (\ref{eq: Ford 1}), or  Eq. (\ref{eq: Ford 2}),
one of the expressions for the energy
difference and then equate it with one of the expressions for the age of the universe, i.e., Eq. (\ref{eq: carmeli}) or Eq. (\ref{eq: macdonald}),
we obtain the values for the scale factor $a$
(for multiple solutions, we select the smallest non-zero value of $a$). 
Then, we estimate the corresponding length, i.e., $L_0a$, the  redshift, i.e., $z$,
and finally the decoherence time, i.e., $t_d$.
\par\noindent
By inspection, we find that for a number of cases, which
in fact exhaust the most important quantum systems,
the decoherence time is significantly less than the age of our Universe, {\it but}
is not less than the Planck time.
Therefore, for these systems, gravitation may indeed have been
the dominant source of decoherence
{\footnote{
{In those cases for which the decoherence time turns out to be smaller than
the Planck time, about $10^{-43}$ s, the equations used here and the subsequent results
cease to remain trustworthy, as it is expected that
a full theory of quantum gravity is needed to study the sub-Planckian regime.
Away from this regime however, the decoherence time formulas
Eqs.(\ref{eq: Blencowe}-\ref{eq: Ford 2})
derived from a minimal set of assumptions
are quite robust, and expected to hold for the situations we consider.
}.}}.
We also ensure that those systems are valid in the high or low temperature approximations
as required by equations (\ref{eq: Blencowe})-(\ref{eq: Ford 2}).
In Section \ref{sec: Discussion}, we discuss  our results and the implications of our calculations.\\
%
\section{Calculations}
%
\subsection{Using the Cosmic Time Formula from Ref. \cite{Carmeli:2005if}}
%
\vspace{-2ex}
\par\noindent
Here we utilize Eq.(\ref{eq: carmeli}) to determine the age of the Universe, for a system undergoing decoherence.
%
%
%
\begin{table}[h!]
\centering
\begin{tabular}{|l|l|l|l|}
\hline
\multicolumn{1}{|c|}{{ $\Delta\omega$ (J)}} & $L_0a$ (m) & $z$ & $t_d$ (s) \\ \hline
UR & $9\times10^{-67}$ & $10^{93}$ & $10^{-168}$ \\ \hline
NR (g) & $4\times10^{-28}$& $2\times10^{54}$  & $2\times10^{-91}$ \\ \hline
NR (e)  & $10^{-30}$& $8\times10^{56}$ & $2\times10^{-96}$ \\ \hline
$1.6\times10^{-19}$ & $0$ & $\infty$ & $0$ \\ \hline
\end{tabular}
\caption{Determining the redshift, i.e., $z$, the size of the Universe, i.e., $L_0a$, and the age of the Universe, i.e.,
$t_d$, for various energies by using Eqs. (\ref{eq: Blencowe}) and (\ref{eq: carmeli}).}
\label{tab: Blencowe}
\end{table}
%
%
%
%
\begin{table}[h!]
\centering
\begin{tabular}{|l|l|l|l|}
\hline
\multicolumn{1}{|c|}{{$\Delta\omega$ (J)}}               & $L_0a$ (m) & $z$       & $t_d$ (s)          \\ \hline
UR                                                        & 0        & $\infty$          & 0                  \\ \hline
NR (g)                                             & $2\times10^{-5}$   & $6\times10^{31}$     & $3\times10^{-46}$ \\ \hline
NR (e)                                             & $3\times10^{-44}$ & $3\times10^{70}$& $10^{-123}$ \\ \hline
$1.6\times10^{-19}$                                   & $3\times10^{25}$    & 34                & $7\times10^{14}$*                \\ \hline
\end{tabular}
\caption{Determining the redshift, i.e., $z$, the size of the Universe, i.e., $L_0a$, and the age of the Universe, i.e., $t_d$,
for various energies by using Eqs. (\ref{eq: Ford 1}) and (\ref{eq: carmeli}).}
\label{tab: Ford 1}
\end{table}
%
%
%
%
\begin{table}[h!]
\centering
\begin{tabular}{|l|l|l|l|l|l|}
\hline
\multicolumn{1}{|c|}{{$\Delta\omega$ (J)}} & $\lambda_g$ (m) & $L_0a$ (m)  & $z$                    & $t_d$ (s)            \\ \hline
UR                                                       & 1            & $10^{-26}$  & $5\times10^{52}$& $3\times10^{-88}$ \\ \hline
                                                       & $L_0a$                    &      $6.2\times10^{25}$         &       13.1     &    $4.4\times10^{15}$*                \\ \hline
NR (g)                                            & 1                  & $3\times10^{8}$  & $2\times10^{18}$& $2\times10^{-19}$*       \\ \hline
                                                        & $L_0a$        & $6.2\times10^{25}$  & $13.3$& $4.3\times10^{15}$*         \\ \hline
NR (e)                                                & 1                & $3\times10^{-17}$  & $3\times10^{43}$& $9\times10^{-70}$     \\ \hline
                                                         & $L_0a$          & $2\times10^{-50}$  & $4\times10^{76}$& $6\times10^{-136}$   \\ \hline
$1.6\times10^{-19}$                                   & 1               & $2\times10^{13}$  & $4\times10^{13}$& $6\times10^{-10}$*  \\ \hline
                                                    & $L_0a$           & $3\times10^{4}$   & $3\times10^{22}$  & $9\times10^{-28}$*  \\ \hline
\end{tabular}
\caption{Determining the redshift, i.e. $z$, the size of the Universe, i.e., $L_0a$, and the age of the Universe, i.e., $t_d$,
by using Eqs.  (\ref{eq: Ford 2}) and (\ref{eq: carmeli}), for various energies, and assuming that the graviton
bath energy density  is equal to the energy density of dark matter.}
\label{tab: Ford 2}
\end{table}
\vspace{-2ex}
\subsection{Using the Cosmic Time Formula from  Ref. \cite{Macdonald:2006sa}}
%
\vspace{-1ex}
\par\noindent
Here we utilize Eq.  (\ref{eq: macdonald}) to determine the age of the Universe, for a system undergoing decoherence.
%
\begin{table}[h!]
\centering
\begin{tabular}{|l|l|l|l|}
\hline
\multicolumn{1}{|c|}{{$\Delta\omega$ (J)}}  & $L_0a$ (m) & $z$& $t_d$ (s) \\ \hline
UR  & $7\times10^{-36}$ & $10^{62}$& $5\times10^{-107}$ \\ \hline
NR (g)  & $0.46$& $2\times10^{27}$ & $6\times10^{-24}$* \\ \hline
NR (e)  & $10^{-22}$ & $7\times10^{48}$& $2\times10^{-80}$ \\ \hline
$1.6\times10^{-19}$ & 0& $\infty$ & $0$ \\ \hline
\end{tabular}
\caption{Determining the redshift, i.e., $z$, the size of the Universe, i.e., $L_0a$, and the age of the Universe, i.e., $t_d$,
for various energies by using Eqs.  (\ref{eq: Blencowe}) and (\ref{eq: macdonald}).}
\label{tab: Blencowe1}
\end{table}
\begin{table}[h!]
\centering
\begin{tabular}{|l|l|l|l|}
\hline
\multicolumn{1}{|c|}{{$\Delta\omega$ (J)}}           & $L_0a$ (m)   & $z$         & $t_d$ (s)          \\ \hline
UR                                                     & $2\times10^{-27}$      & $4\times10^{63}$               & $2\times10^{-78}$                  \\ \hline
NR (g)                                           & $9\times10^{4}$    & $9\times10^{21}$      & $6\times10^{-16}$* \\ \hline
NR (e)                                             & $7\times10^{-21}$& $10^{47}$ & $10^{-53}$ \\ \hline
$1.6\times10^{-19}$                            & $2\times10^{24}$           & $473$    &    $5\times10^{13}$*          \\ \hline
\end{tabular}
\caption{Determining the redshift, i.e., $z$, the size of the Universe, i.e., $L_0a$, and the age of the Universe, i.e., $t_d$,
for various energies using Eqs.  (\ref{eq: Ford 1}) and (\ref{eq: macdonald}).}
\label{tab: Ford 1.1}
\end{table}
\begin{table}[h!]
\centering
\begin{tabular}{|l|l|l|l|l|l|}
\hline
\multicolumn{1}{|c|}{{$\Delta\omega$ (J)}}  & $\lambda_g$ (m)          & $L_0a$ (m)  & $z$           & $t_d$ (s)            \\ \hline
UR                                             & 1                & $2.3$ & $4\times10^{26}$& $7\times10^{-23}$*  \\ \hline
                                                   & $L_0a$                         &   $2\times10^{27}$  &  $-0.59$       &   $9\times10^{17}$             \\ \hline
NR (g)                                             &    1             &        $10^{13}$      &    $7\times10^{13}$          &    $9\times10^{-4}$*           \\ \hline
                                                     &  $L_0a$                  & $10^{26}$    & $5.1$      & $4\times10^{16}$*  \\ \hline
   NR (e)                                               & 1             & $2\times10^{-6}$   & $5\times10^{32}$ & $5\times10^{-32}$*  \\ \hline
                                                         & $L_0a$         & $3\times10^{-12}$  & $3\times10^{38}$ & $10^{-40}$* \\ \hline
 
$1.6\times10^{-19}$                                    & 1              & $10^{32}$      & $-1$        & $7\times10^{18}$  \\ \hline
                                                     & $L_0a$           & $3\times10^{-7}$ & $3\times10^{33}$    & $3\times10^{-33}$*  \\ \hline
 
\end{tabular}
\caption{Determining the redshift, i.e., $z$, the size of the Universe, i.e.,  $L_0a$, and the age of the Universe, i.e., $t_d$,
by using Eqs.  (\ref{eq: Ford 2}) and (\ref{eq: macdonald}), for various energies, and assuming that the graviton bath
energy density is equal to the energy density of dark matter.}
\label{tab: Ford 2.1}
\end{table}

\section{Discussion}
\label{sec: Discussion}
%
%
%
%
\par\noindent
%
%
%
 
We note from Tables I-VI that the decoherence times marked by an asterisk (${}^\ast$) are less than the age of the Universe,
{\it but} not smaller than the Planck time
(as noted earlier for the latter, a full theory of quantum gravity would be needed, and
the equations used here cease to remain trustworthy).
These are the ultra-relativistic particles, non-relativistic gravitons, non-relativistic electrons and
atomic systems. While these may not form an exhaustive list, they cover much of the spectrum of quantum systems.

Several points are in order.
First, we note that Eq. (\ref{eq: Blencowe}) requires the system to be macroscopic, while neither Eq.
(\ref{eq: Ford 1}) nor Eq. (\ref{eq: Ford 2}) have constraints regarding the size of the system.
Furthermore, while Eq.  (\ref{eq: Blencowe}) applies to high temperature systems, Eqs.  (\ref{eq: Ford 1}) and (\ref{eq: Ford 2})
are valid for systems at low temperatures.
Second, the decoherence time for a macroscopic system composed of 
non-relativistic gravitons, as calculated
using Eqs. (\ref{eq: Blencowe}) and (\ref{eq: carmeli}), lies between the GUT and Hadron epochs\footnote{The GUT
epoch starts at time $10^{-43}s$ while the Hadron epoch starts at time $10^{-6}s$ \cite{Turner:2009zza,quarks}.}.

\par\indent
Third, it is also noteworthy that Eqs.  (\ref{eq: Ford 1}) and (\ref{eq: Ford 2}) are calculated in
the low-temperature limit \cite{DeLorenci:2014vwa}. Therefore, for the systems that appear in Tables II and V, we have to
discard the results for which the Universe appears to be too hot. In the last case, with $\Delta\omega=1$ eV,
the age of the Universe according to Eq.  (\ref{eq: carmeli}) is approximately 22.2 million years, while  Eq.(\ref{eq: macdonald}) suggests an age of about 1.6 million years.
This indicates that a system whose difference in energy levels is around 1 eV would
decohere during the Dark Ages\footnote{The Dark Ages of the Universe are considered to be between $3.8\times10^{5}$ years
and   $150\times10^{6}$ years after the Big Bang \cite{Turner:2009zza, quarks}.}.
\par\indent
Fourth, in Tables \ref{tab: Ford 2} and \ref{tab: Ford 2.1}, it is evident that the decoherence time depends
on the characteristic wavelength of the graviton. The characteristic wavelengths of
$\lambda_g=1$ and $\lambda_g=L_0a$ were chosen to demonstrate two possible extremes: a small constant wavelength and a wavelength that grows at the same rate as the Universe.
\par\indent
It is usually assumed that gravitons
have the  speed of light, i.e., $c$. For subluminal gravitons, Eq.  (\ref{eq: Ford 2}) can still be utilized,
though the graviton's  speed, $v_g$, has to be incorporated probably by modifying $\lambda_g$, such that it becomes
$\lambda_g=\lambda_g(v_g)$.
It should also be stressed that the characteristic wavelength of the graviton has to be smaller than the size of the Universe.
Thus, we can eliminate one of the cases in which $\lambda_g=1$. 
As already mentioned, Eq.  (\ref{eq: Ford 2}) is computed in the low-temperature limit
\cite{DeLorenci:2014vwa}. Thus, two of the only remaining cases  correspond to $z=13.1$ and $z=13.3$, with the age of the Universe to be around 140 million years. Therefore, decoherence would occur near the end of the Dark Ages. We also have $z=3\times10^{38}$ and $z=3\times10^{33}$, with a decoherence time of $3\times10^{-33}$ s and $10^{-40}$ s, respectively, occurring between the GUT and Hadron epochs.
The final remaining case is $z=5.1$, at an age of 1.27 billion years.
In this case, decoherence would occur  almost at the start of Galaxy formation and evolution epoch\footnote{Galaxy
formation and evolution epoch took place between $10^{9}$ years and $10\times10^{9}$ years \cite{Turner:2009zza, quarks}.}.
\par\indent
Fifth, as already mentioned, the derivations of the formulas for decoherence time involve either the
high-temperature approximation or the low-temperature approximation.
Thus, it is anticipated that a new formula for decoherence time will be  more accurate both at high and low temperatures.
This will ensure that  decoherence time could be calculated for a particle or a
system transitioning from a high-temperature environment (the early Universe) to a low-temperature environment (current time).
\par\indent
Sixth,  decoherence due to scattering by various sources was considered in Ref. \cite{Tegmark:1993yn} and
the author demonstrated that the decoherence time varied depending on the size of the system.
We might expect similar results with decoherence due to gravitons. In addition, for a system whose $\Delta\omega$ is
equivalent to that of a non-relativistic particle,  the decoherence time depends on the particle's mass as
given in Eq. (\ref{energy3}). Moreover, we might expect this dependence of decoherence time on particle's mass to
hold also for the case of ultra-relativistic time but in this case the dependence will not come through $\Delta\omega$.
Thus, a new theoretical model is expected to incorporate not only
the superposition of energy levels of a system, but also the system's size and mass \cite{Zurek:2003zz}.
Otherwise, two macroscopic systems
whose sizes are radically different (for example, a planet and an asteroid) would have the same decoherence time,
as long as $\Delta\omega$ is the same. This result seems to make little physical sense.
Furthermore, as  the Universe evolves, it initially includes particle-sized objects before eventually
forming macroscopic systems. As these particle-sized objects are initially relativistic, it will be necessary in the new formula
for the decoherence time in the early Universe to account for transitioning from relativistic speeds to non-relativistic speeds,
changing the energy level, and changing the mass of the system.\\
\indent
Equally important, Eqs. (\ref{eq: carmeli})
and (\ref{eq: macdonald}) both describe the age of the Universe,  however
they have different ranges of validity. By utilizing the accepted values for the redshift of each epoch, it can be shown
that Eq.  (\ref{eq: carmeli}) yields a closer estimate to the accepted age of the  Universe from GUT epoch  till 
the start of Nucleosynthesis Epoch as well as from the start of the Reionization epoch until the present time.
Eq. (\ref{eq: macdonald}) yields a closer estimate to the accepted age of the Universe  from the end of
Recombination Epoch  until the present time.
 
\par\indent
Finally, further research is needed to determine whether gravitational decoherence is the dominant form of decoherence
in the early Universe, or if it is necessary to consider multiple sources of decoherence. Assuming decoherence is induced
solely by gravitation, one interpretation of our calculations is that gravitational decoherence may be insufficient
in some situations to produce a classical Universe. In each of the cases considered here, the decoherence time was of the order
of millions or billions of years. Such a timescale is heavily dependent upon the system and environment under consideration.
This may alternatively indicate that different sectors of the Universe decohere before others. \\
%
\subsection{Acknowledgements}
%
\vspace{-2ex}
\par\noindent
We thank P. Bosso for useful discussions.
SD and MPGR would like to  thank the Natural Sciences and Engineering Research Council of Canada and the University of Lethbridge for support.
We thank the anonymous Referee for useful comments which has helped in
improving the paper.
 
%

 
\end{document}